# Effect of MTU length on child-adult difference in neuromuscular fatigue

**Authors:** Enzo Piponnier[1], Vincent Martin[1], Emeric Chalchat[1], Bastien Bontemps[1], Valérie Julian[2], Olivia Bocock[2], Martine Duclos[2], Sébastien Ratel[1]

**Authors affiliations:** [1] Clermont-Auvergne University, AME2P, F-63000 Clermont-Ferrand, France, [2] Clermont University Hospital, Clermont-Ferrand, France.

**Corresponding author:**

Dr. Sébastien RATEL

UFR STAPS - Laboratoire AME2P

Campus Universitaire des Cézeaux

3 rue de la Chébarde

TSA 30104 - CS 60026

63178 AUBIERE CEDEX

FRANCE

Tel: + 33 (0)4 73 40 54 86

Fax: +33 (0)4 73 40 74 46

E-mail: sebastien.ratel@uca.fr

# ABSTRACT

**Purpose**: The purpose of this study was to compare the development and etiology of neuromuscular fatigue of the knee extensor muscles (KE) at different muscle-tendon unit (MTU) lengths during repeated maximal voluntary isometric contractions (MVIC) between boys and men.

**Methods**: Twenty-two pre-pubertal boys (9-11 years) and 22 men (18-30 years) performed three KE fatigue protocols at short (SHORT), optimal (OPT) and long (LONG) MTU lengths, consisting of repeating 5-s MVIC interspersed with 5-s passive recovery periods until torque reached 60% of the initial MVIC torque. The etiology of neuromuscular fatigue was identified using non-invasive methods such as surface electromyography, near-infrared spectroscopy, magnetic nerve stimulation and twitch interpolation technique.

**Results**: The number of repetitions was significantly lower in men at OPT (14.8±3.2) and LONG (15.8±5.8) than boys (39.7±18.4 and 29.5 ±10.2, respectively; p<0.001), while no difference was found at SHORT between both age groups (boys: 33.7±15.4, men: 40.9±14.2). At OPT and LONG boys showed a lower reduction in the single potentiated twitch ($Qtw_{pot}$) and a greater decrease in the voluntary activation level (VA) than men. At SHORT, both populations displayed a moderate $Qtw_{pot}$ decrement and a significant VA reduction (p<0.001). The differences in maximal torque between boys and men were almost twice greater at OPT (223.9 N.m) than at SHORT (123.3 N.m) and LONG (136.5 N.m).

**Conclusion**: The differences in neuromuscular fatigue between children and adults are dependent on MTU length. Differences in maximal torque could underpin differences in neuromuscular fatigue between children and adults at OPT and SHORT. However, at LONG these differences do not seem to be explained by differences in maximal torque. The origins of this specific effect of MTU length remain to be determined.

# INTRODUCTION

Neuromuscular fatigue is defined as any change that occurs in the central nervous system and/or muscles from exercise, inducing a lower muscular torque level than expected during voluntary or evoked contractions (1). Performance fatigability, defined as the time required to reach a given level of exhaustion (e.g. time to task failure), was found to be lesser in children than adults during repeated or sustained maximal voluntary isometric contractions (MVIC) (2, 3) or during whole-body dynamic activities (4, 5). However, performance fatigability was found to be similar in children and adults during submaximal voluntary contractions (6, 7) or throughout MVIC of plantar flexor (PF) muscles (8). Different factors could underpin these differences in the development and etiology of neuromuscular fatigue between children and adults such as the muscle groups investigated (8), muscle phenotype (9), muscle activation (10) or torque level (11).

The torque level has a significant influence on neuromuscular fatigue. Indeed, it has been shown that the differences in neuromuscular fatigue between men and women were significant when men developed a higher torque level (12), while the differences were abolished when both developed a similar torque level (13). Indirect evidence supporting this proposal between children and adults was also provided by Ratel et al. (11). These authors showed that the development and etiology of neuromuscular fatigue were not different when the initial torque level was used as a co-variable. In addition, Piponnier et al. (8) have recently reported that the development and etiology of neuromuscular fatigue were comparable between children and adults with the PF but not with the knee extensor (KE) muscles during an intermittent fatigue protocol. Interestingly, in this study, the difference in absolute maximal torque between boys and men before the fatigue protocol was almost twice-lower on the PF than the KE muscles (107.4 vs. 219.1 N.m, respectively). Hence, these authors hypothesized that the reduced difference in torque level may have contributed to reduce the difference in the development and etiology of neuromuscular fatigue between boys and men with the PF muscles. However, scientific evidence showing the influence of maximal torque on differences in neuromuscular fatigue between children and adults is still lacking since any child-adult comparison of neuromuscular fatigue at the same absolute force is strongly biased. Matching maximal torque levels between healthy adults and children is impossible, and matching submaximal torque levels would not represent the same relative effort since healthy children and adults do not produce comparable maximal torque levels.

An alternative suggested by Piponnier et al. (8) would be to manipulate the muscle-tendon unit (MTU) length to examine the effect of the difference in torque level between children and adults on the difference in neuromuscular fatigue between both populations. Indeed, MTU length can influence the torque level (14) by modifying actin-myosin cross-bridge number (15), agonist and antagonist muscles activation (16), moment arm (17) and passive torque (18). The torque level decreases at short MTU length because the actin-myosin cross-bridge number and activation level decrease in this condition. Although still debated in the literature (16, 19), the torque depression could be partly compensated by the co-activation reduction at short MTU length (15, 16). The decrement in torque level at short MTU length is also explained by a reduction of the moment arm (related to joint angle variation) and passive torque (17, 18).

MTU length can also influence the extent and origin of neuromuscular fatigue. It has been reported at short MTU length that adults display lesser performance fatigability (20, 21), lesser peripheral fatigue (22, 23) and greater central fatigue (21) than at optimal and long MTU lengths. The lesser performance fatigability and peripheral fatigue at short MTU length could be partly explained by a lower torque level than at optimal length, where greater alterations in peripheral mechanisms [e.g. excitation-contraction (E-C) coupling] are observed (22, 23). This lower alteration in peripheral mechanisms could be also related to lower energy cost, as measured by near-infrared spectroscopy (NIRS) (24) at short than at optimal MTU length. Lichtwark and Barclay proposed that the reduced energy cost in fatigued state may be attributed to a lower tendon stiffness (25). Interestingly, these authors also suggested that a compliant tendon may act as a mechanical buffer, that could additionally protect from the development of peripheral fatigue, and therefore from the reduction of maximal force. Finally, the longer exercise duration at short MTU length could account for the greater central fatigue (23).

Variation of MTU length could affect the differences in neuromuscular fatigue between children and adults. Firstly, it could modulate the difference in torque level of the KE muscles between children and adults (26). It has been shown that at short KE MTU length (20° of flexion; 0° = full extension), the difference in torque level between children and adults (~ 50 N.m) is four times lower than at a longer MTU length (90°; ~ 200 N.m). Secondly, at short MTU length, differences in MTU mechanical properties between children and adults could be reduced since (i) children have a lower MTU stiffness than adults (Kubo et al 2014) and (ii) MTU stiffness is lower at short than at long MTU length in adults (Kubo et al 2006). Taken together, differences in magnitude and etiology of neuromuscular fatigue between children and adults could therefore be reduced at short MTU length. However, scientific evidence is still lacking to support this assumption.

Therefore, the purpose of the present study was to compare child-adult differences in the development and etiology of neuromuscular fatigue at different KE MTU lengths. We hypothesized that at short MTU length (i.e. reduced child-adult differences in torque level), children and adults could display reduced performance fatigability, mainly accounted for by central fatigue, rather than peripheral fatigue, compared to long and optimal MTU lengths.

## MATERIAL AND METHODS

### Participants

Twenty-two pre-pubertal boys (9-11 years) and 22 men (18-30 years) volunteered to participate in the present study. All the participants were involved in different physical activities such as rugby, soccer, judo, etc. To be included, all participants had to spend less than 4 h per week in recreational physical activity and were free of any medical contra-indication to physical activity. The local ethics committee (Protection Committee of People for Biomedical Research South-East 6; authorization number, AU 1268) approved the present study. All participants were fully informed of the experimental procedures and gave their written consent/assent before any testing was conducted. The written consent of the parents/guardians was also obtained for the children.

### Experimental procedure (design)

All participants achieved four experimental sessions separated by at least one week. During the first session, participants' physical characteristics (anthropometric measurements, maturation status) were collected; a medical practitioner (pediatrician for the children) performed a clinical examination. The participants were also familiarized with the experimental procedures. Furthermore, at the end of this session, participants performed a series of MVIC of the KE in order to determine their optimal knee angle for maximal torque production. They had to maximally contract their muscles at different knee angles (30°, 50°, 70°, 75°, 80°, 85°, 90°, 100°; 0° = full extension) in a randomized order. During the three following sessions, participants performed an intermittent voluntary fatigue protocol (see below for further details) at short (SHORT; 30°-knee angle), optimal (OPT; 77.0 ± 5.5 and 75.2 ± 5.0° in boys and men, respectively) and long (LONG; 110°) KE MTU lengths.

At the beginning of the three sessions, participants were equipped and performed a progressive warm-up (4 contractions up to ~50% MVIC, 4 contractions up to ~80% MVIC and 2 contractions up to 100% MVIC with a rest of 30-s between each contraction). Then, they performed three MVIC of the KE and two MVIC of the knee flexors with a 2-min rest between each contraction. Before participants performed the intermittent voluntary fatigue test, a 5-min rest period was allowed in order to prevent any extensive fatigue. The intermittent voluntary fatigue protocol consisted in a repetition of 5-s MVIC of the KE interspersed with 5-s passive recovery periods until the voluntary torque reached the target value of 60% of its initial value over three consecutive MVIC. The participants were not informed of this criterion of task failure and had no visual feedback of torque output throughout the exercise. The investigators strongly encouraged the participants during each maximal effort throughout the experimental protocol. The number of repetitions was considered as the criteria to quantify performance fatigability.

**Anthropometric measurements and maturation assessment**

A digital weight scale (TANITA, BC-545N, Japan) was used to measure body mass to the nearest 0.1 kg and barefoot standing height was assessed to the nearest 0.1 cm with a wall-mounted stadiometer (TANITA, HR001, Japan). Body mass index (BMI) was calculated as body mass (kg) divided by height squared (m²).

We used two methods to assess boys' maturation status. First, Tanner stages were determined from self-reported assessment on the basis of pubic hair and testicular/penis development (27), the children being assisted by their parents while completing the questionnaire. Second, age at peak height velocity (APHV) was determined to assess somatic maturity and calculated using chronological age, height and sitting height of the boys. Its calculation was based on sex-specific regression equations according to the method proposed by Mirwald et al. (28).

**Torque measurement**

Voluntary and evoked torques were measured using dynamometer (Biodex System 3, Biodex, Shirley, NY). Volunteers were comfortably positioned on an adjustable chair with the hip joint flexed at 60° (0° = neutral position). The participant position was standardized and was the same during the four sessions. The knee joint was set according to the considered session (SHORT, OPT, LONG). The axis of rotation of the dynamometer was aligned with the lateral femoral condyle of the right femur and the lever arm was attached 1-2 cm above the lateral malleolus with a Velcro strap. During each contraction, the participants were instructed to grip the lateral handles in order to further stabilize the pelvis. Torque data were corrected for gravity, digitized and exported at a rate of 2 kHz to an external analog-to-digital converter (PowerLab 8/35; ADInstrument, New South Wales, Australia) driven by the LabChart 7.3 Pro software (ADInstrument, New South Wales, Australia).

**EMG recordings**

In order to achieve low impedance at the skin-electrode interface (Z < 5 kΩ), skin was prepared prior to the EMG surface electrodes placement by shaving, lightly abrading with sandpaper and cleaning with alcohol. EMG electrodes (Ag-AgCl, Blue Sensor N-00-S, Ambu, Denmark) were positioned on the right muscle bellies of the *vastus lateralis* (VL), *vastus medialis* (VM), *rectus femoris* (RF), *biceps femoris* (BF), according to the SENIAM (Surface Electromyography for Non-Invasive Assessment of Muscles) recommendations (29), with an inter-electrode distance of 20 mm. EMG signals were amplified (Dual BioAmp, ML 135, ADInstruments, New South

Wales, Australia) within a bandwidth frequency ranging from 10 to 500 Hz (common mode rejection ratio > 85dB, gain = 1000) and simultaneously digitized with torque signal by using the external analog-to-digital converter driven by the LabChart 7.3 Pro software. EMG signals were sampled at a frequency of 2 kHz during voluntary and evoked contractions.

**Magnetic nerve stimulation**

We used a 70-mm figure-of-eight coil connected to two Magstim $200^2$ stimulators linked by the Bistim$^2$ module (peak magnetic field strength 2.5 T, stimulus duration 115 μs; Magstim, Witland, Dyfed, UK) to stimulate the femoral nerve and induce involuntary KE muscles contractions. The coil was placed high in the femoral triangle in regard of the femoral nerve. Small spatial adjustments were initially performed in order to determine the optimal position where the greatest unpotentiated twitch amplitude and the greatest compound muscle action potentials (M-wave) were evoked. Prior to the testing procedure, the optimal stimulation intensity ($I_{opt}$) (i.e. the intensity where unpotentiated twitch and concomitant M-waves amplitudes of the VL reached its maximal value and started to plateau) was determined from recruitment curves. During subsequent testing procedures, participants were stimulated at supra-maximal intensity (100% of the magnetic stimulator output) in order to overcome the potential confounding effect of the axonal hyperpolarization (30). This intensity corresponded to 106.9 ± 8.6%, 109.2 ± 11.1% and 104.6 ± 6.4 of the $I_{opt}$ in boys during SHORT, OPT and LONG sessions, respectively. The corresponding values for the men were 107.6 ± 12.7%, 111.9 ± 16.5% and 105.6 ± 10.5% of the $I_{opt}$. These supra-maximal intensities were statistically higher than the $I_{opt}$ ($p < 0.001$) and were not statistically different between groups.

**Muscle oxygenation**

Muscle oxygenation in the VL was continuously recorded 5 min before and during the entire intermittent fatigue protocol by means of a three-channel, portable continuous-wave NIRS device (PORTAMON, peak wavelengths of 750 nm and 850 nm, Artinis Medical System, Zetten, The Netherlands). Oxygenated, deoxygenated, and total hemoglobin concentration, expressed in micromolars, were calculated from changes in optical density by using a modified Lambert-Beer law for which a differential path length factor is used to correct photon scattering within the tissue (31). The tissue saturation index (TSI), expressed in percent, was also calculated. The TSI corresponds to the oxygenated hemoglobin proportion of total hemoglobin and is derived from the relative absorption coefficients obtained from the slopes of light attenuation at three interoptode distances and by taking the diffusion scattering law into account (32). The NIRS probe was positioned on the VL midway between the lateral epicondyle and the great trochanter of the femur. It was securely strapped to ensure that the probe did not move during the experimental session. An opaque black fabric was placed and fixed over the probe to prevent signal interference by ambient light. The subcutaneous fat layer thicknesses, where the probe was placed, were measured using a B-mode ultrasound (Echo Blaster 128 CEXT-1Z; Telemed, Vilnius, Lithuania) with a 7.5 MHz linear array transducer. Adipose tissue thicknesses over the VL muscle were 6.6 ± 2.5 and 5.8 ± 2.3 mm in boys and men, respectively (P < 0.31). Considering that the adipose tissue thickness was relatively low and the penetration depth of the NIRS signal is more or less half of the emitter-detector distance (4 cm), the changes in signals reflected the muscle hemodynamic changes. NIRS signals were recorded and sampled at 10 Hz using the Oxysoft software (Artinis Medical System, Zetten, The Netherlands).

**Data analysis**

**Peripheral fatigue indicators.** To examine the time course of peripheral fatigue, potentiated single twitches ($Qtw_{pot}$), evoked at supra-maximal intensity, were measured before, every five MVIC and after the last MVIC of the intermittent fatigue protocol by stimulating the femoral nerve 3-s after the cessation of the current MVIC. Concomitant peak-to-peak M-wave amplitudes ($M_{max}$) were also measured on VL, VM and RF in order to assess sarcolemma excitability. Before and after each fatigue protocol, doublets at 10Hz and 100Hz ($Dt_{10Hz}$ and $Dt_{100Hz}$) were evoked. $Dt_{100Hz}$ is usually considered as an indicator of muscle contractile properties. Intensity of double stimulations was set to 60% of the maximal magnetic stimulator output since higher intensities were painful especially in boys. The $Dt_{10Hz}$-to-$Dt_{100Hz}$ ratio ($Dt_{10Hz}/Dt_{100Hz}$) obtained from magnetic stimulation was calculated and used to assess changes in E-C coupling (33).

To quantify muscle oxygenation, TSI signals were expressed as the magnitude of change from the baseline (i.e. the mean value over 30 s before the onset of the first contraction). For all the contractions performed, the minimum TSI value reached during the contraction (ΔTSI) was measured. Important ΔTSI decreases indicate greater $O_2$ demand relative to $O_2$ supply (34).

**Central fatigue indicators.** Twitch interpolation technique was used to determine VA. Superimposed single twitch ($Qtw_s$) was evoked at supra-maximal intensity during MVIC after the torque had reached a plateau. Then, $Qtw_s$ and $Qtw_{pot}$ were used to quantify VA before, every five MVIC and during the last MVIC as proposed by Merton (35):

$$VA\,(\%) = \left[1 - \left(Qtw_s \cdot Qtw_{pot}^{-1}\right)\right] \cdot 100$$

The root mean square (RMS) values of the instrumented muscles were calculated during MVIC over a 0.5-s period after the torque had reached a plateau and before the superimposed stimulation was evoked. These RMS values were normalized to respective $M_{max}$ ($RMS.M_{max}^{-1}$) to account for differences in muscle mass and potential changes/differences in sarcolemmal excitability. Finally, the level of antagonist co-activation (%CoAct) was determined every five MVIC by computing the RMS value of BF during intermittent contractions ($RMS_{anta}$) and the RMS value of BF during knee flexion MVIC, recorded before the fatigue protocol ($RMS_{anta-max}$), as follows:

$$\%CoAct = \left(RMS_{anta} \cdot RMS_{anta-max}^{-1}\right) \cdot 100$$

**Statistical analysis**

All variables measured during the intermittent fatigue protocols were linearly interpolated between the nearest values at 20%, 40%, 60% and 80% of number of repetitions (%REP) in order to fairly compare the age groups (children vs. adults) and MTU lengths (SHORT vs. OPT vs. LONG). Values at 0%REP and 100%REP corresponded to pre- and post-fatigue values, respectively.

Data were screened for normality of distribution and homogeneity of variances using Shapiro-Wilk normality test and the Bartlett test, respectively. The total number of repetitions was compared between age groups and MTU lengths using a two-way ANOVA (age group × MTU length). Differences in absolute values and relative changes to the initial values of voluntary and evoked torque, EMG, NIRS and VA variables were analyzed using a three-way (age group × MTU length × %REP) ANOVA with repeated measures. When ANOVA revealed significant effects or interactions between factors, a Tukey HSD *post hoc* test was applied to test the differences between means. The size effect and statistical power were also calculated when significant main or interaction effects were detected. The size effect was assessed using the partial eta-squared ($\eta^2$) and ranked as follows: ~ 0.01= small effect, ~ 0.06 = moderate effect, ≥ 0.14 = large effect. Linear regression models were used to determine correlations between initial absolute torque and relative $Qtw_{pot}$ or VA changes over the course of the fatigue protocol. Data from the two age groups and the three muscle lengths were pooled to determine these correlations. Statistical tests were performed using the Statistica 8.0 sofware (StatSoft, Inc,

USA). Data are reported as mean ± standard deviation (SD). The α–level for statistical significance was set at $p < 0.05$.

# RESULTS

## Participants' characteristics

Participants' physical characteristics are described in Table 1. As expected, the boys displayed significantly lower values for height, body mass and BMI compared with men. All the boys were pre-pubertal.

*Table 1: Age and physical characteristics of the two groups. \*\*\*: significantly different from boys at $p < 0.001$.*

|  | Boys (n=22) | Men (n=22) |
|---|---|---|
| **Age (y)** | 10.3 ± 0.7 | 21.6 ± 3.3*** |
| **Height (m)** | 1.39 ± 0.07 | 1.78 ± 0.07*** |
| **Body mass (kg)** | 33.4 ± 5.6 | 71.6 ± 8.3*** |
| **BMI (kg.m$^{-2}$)** | 17.1 ± 1.7 | 22.5 ± 2.1*** |
| **Tanner stage (pubic hair)** | I-II | |
| **Tanner stage (testicular size)** | I-II | |
| **APHV (y)** | 13.4 ± 0.4 | |
| **Years to APHV (y)** | -3.3 ± 0.6 | |

*APHV: Age at Peak Height Velocity; BMI: body mass index.*

## MVIC torque

ANOVA revealed a significant interaction effect (age group × MTU length × %REP) for absolute MVIC torque [$F(10;420) = 17.2$; $p < 0.001$, $\eta^2 = 0.29$, power = 1.0]. As expected, men displayed significantly higher absolute MVIC torque values than boys at any MTU length (Fig. 1A). Boys-men initial torque level difference were 123.3, 223.9 and 136.5 N.m at SHORT, OPT and LONG, respectively. In addition, initial absolute MVIC torque at OPT was higher than at LONG ($p < 0.001$) and SHORT in boys and men ($p < 0.001$). A significant age group × %REP interaction effect was found for MVIC torque when expressed as percentage of its initial value [$F(4;168) = 24.7$; $p < 0.001$, $\eta^2 = 0.37$, power = 1.0]. Boys showed a greater decrement in relative MVIC torque than men at 20 and 40%REP, whatever the MTU length (Fig. 1B).

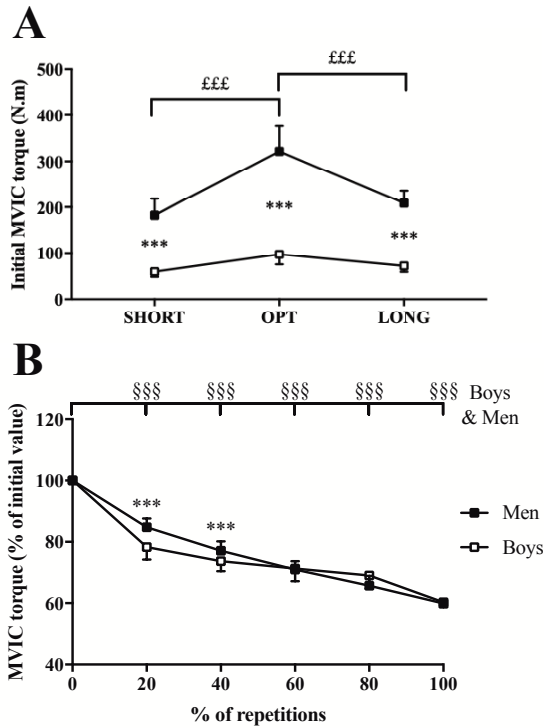

*Figure 1: (A) Initial absolute MVIC torque at short (SHORT), optimal (OPT) and long (LONG) MTU length in boys and men. (B) Time course of relative MVIC torque (the three MTU lengths pooled) during fatigue protocols in boys and men. \*\*\*: significant difference between groups at p < 0.001; £££: significant difference between MTU lengths at p < 0.001; §§§: significantly different from the initial value at p < 0.001.*

**Number of repetitions**

ANOVA showed a significant age group × MTU length interaction effect for the total number of repetitions [$F(2;84) = 28.4$; $p < 0.001$, $\eta^2 = 0.41$, power = 1.0]. Boys achieved more repetitions to reach 60% of their initial MVIC at OPT (39.7 ± 18.4) and LONG (29.5 ± 10.2) than men (14.8 ± 3.2 and 15.8 ± 5.8, respectively; $p < 0.001$). No difference in the number of repetitions was observed at SHORT between boys and men (33.7 ± 15.4 and 40.9 ± 14.2 repetitions, respectively). Men performed a greater number of repetitions at SHORT than OPT and LONG, while *post-hoc* analysis only revealed a difference between OPT and LONG in boys.

**Peripheral fatigue**

***Potentiated twitch torque.*** ANOVA revealed a significant interaction effect (age group × MTU length × %REP) for absolute $Qtw_{pot}$ values [$F(10;420) = 36.4$, $p < 0.001$, $\eta^2 = 0.48$, power = 1.0]. $Qtw_{pot}$ decreased in men at the three MTU lengths (Fig. 2), whereas it significantly decreased in boys only at LONG. A significant age group × MTU length × %REP interaction effect was found for $Qtw_{pot}$, expressed as percentage of its initial value [$F(8;336) = 10.2$; $p < 0.001$, $\eta^2 = 0.20$, power = 1.0]. Men displayed a lower relative $Qtw_{pot}$ decrement at SHORT (-9.3 ± 15.4%) than at OPT (-50.7 ± 14.6%, $p < 0.001$) and LONG (-56.2 ± 9.6%, $p < 0.001$). At LONG, men showed a greater decrement in relative $Qtw_{pot}$ than boys (-22.5 ± 15.7%; Fig. 2C).

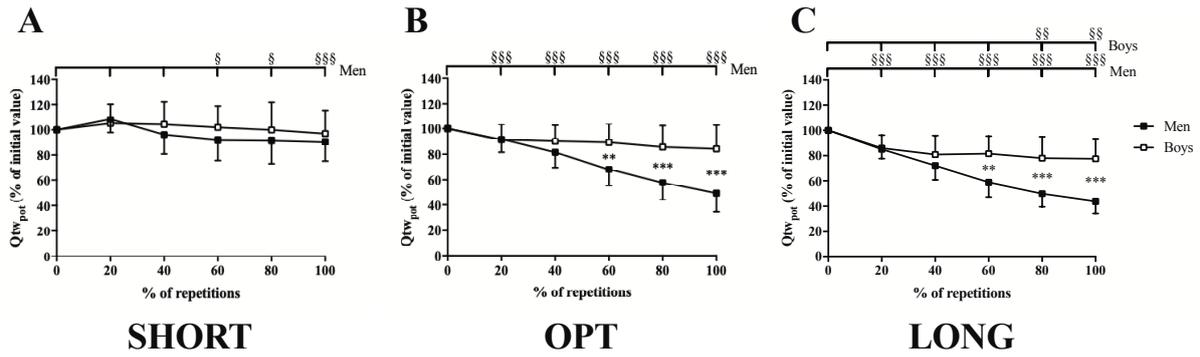

*Figure 2*: Time course of potentiated twitch torque amplitude ($Qtw_{pot}$) at (A) short (SHORT), (B) optimal (OPT) and (C) long (LONG) MTU lengths during fatigue protocols in boys and men. §, §§ and §§§: significantly different from the initial value at $p < 0.05$, $p < 0.01$ and $p < 0.001$, respectively; ** and ***: significant difference between boys and men at $p < 0.01$ and $p < 0.001$, respectively.

Overall, a significant inverse correlation was found between relative $Qtw_{pot}$ changes and initial absolute torque level ($R^2 = 0.37$, $p < 0.05$): the stronger the individual, the greater the $Qtw_{pot}$ decrease during exercise.

***Doublet torque.*** A significant interaction effect (age group × MTU length × %REP) was found for absolute $Dt_{100Hz}$ values [$F(2;126) = 8.7$; $p < 0.001$, $\eta^2 = 0.23$, power = 0.96]. $Dt_{100Hz}$ decreased only in men at OPT (-34.9 ± 21.2%; $p < 0.001$) and LONG (-40.9 ± 17.8%; $p < 0.001$). No significant decrement was observed in boys at SHORT, OPT and LONG.

Furthermore, ANOVA only showed a significant %REP effect [$F(1;42) = 48.2$; $p < 0.01$, $\eta^2 = 0.62$, power = 0.99] for absolute $Dt_{10Hz}/Dt_{100Hz}$ ratio values. $Dt_{10Hz}/Dt_{100Hz}$ ratio significantly decreased following the fatigue protocol whatever the age group and the MTU length ($p < 0.01$) (see Figure, SDC 1, $Dt_{10Hz}/Dt_{100Hz}$ before and after the fatigue protocols at SHORT, OPT and LONG MTU lengths in boys and men). No other main or interaction effect was found for the variation of the $Dt_{10Hz}/Dt_{100Hz}$ ratio between pre- and post-fatigue conditions.

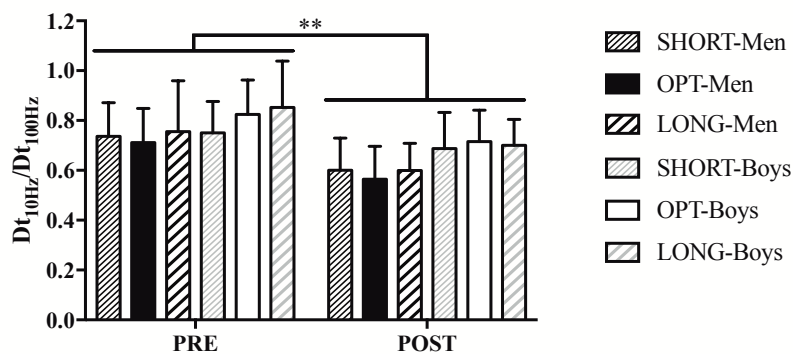

SDC1: Low- to high-frequency torque ratio ($Dt_{10Hz}/Dt_{100Hz}$) before (PRE) and after (POST) the fatigue protocols at short (SHORT), optimal (OPT) and long (LONG) muscle-tendon unit lengths in boys and men. **: significantly different at $p < 0.01$.

***M wave.*** No significant interaction effect was found for absolute VL, VM and RF $M_{max}$ values. ANOVA revealed an age group effect for absolute VM $M_{max}$ values [$F(1;42) = 11.0$; $p < 0.01$, $\eta^2 = 0.23$, power = 0.89]. Boys displayed significantly lower VM $M_{max}$ values than men at all MTU lengths ($p < 0.01$).

*Muscle oxygenation.* ANOVA revealed a significant age group × %REP interaction effect for ΔTSI [F(4;168) = 11.0; p < 0.01, η² = 0.23, power = 0.89]. ΔTSI increased in men between 20%REP and 100%REP, while it remained unchanged throughout the fatigue protocols in boys, whatever the MTU lengths considered (Fig. 3). On average, boys showed a lesser TSI decrement than men during the fatigue protocols whatever the MTU length considered.

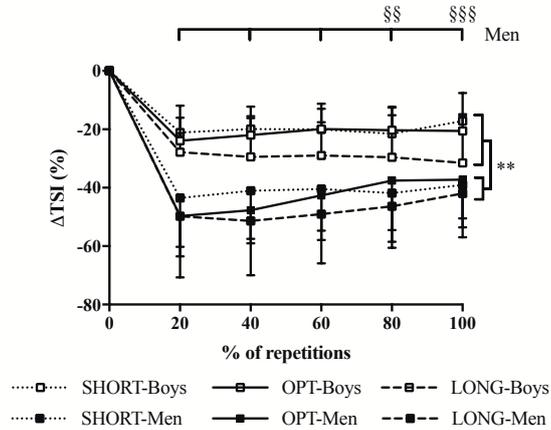

*Figure 3*: Time course of tissue saturation index variation (ΔTSI) of the vastus lateralis during fatigue protocols with the knee extensors at short (SHORT), optimal (OPT) and long (LONG) MTU lengths in boys and men. **: significant difference between boys and men over the fatigue protocol at p < 0.01; §§ and §§§: significantly different from the 20%REP value at p < 0.01 and p < 0.001, respectively.

**Central fatigue**

*Voluntary activation level.* ANOVA revealed an interaction effect (age group × MTU length × %REP) for absolute VA values [F(10;420) = 6.8; p < 0.001, η² = 0.15, power = 0.99]. The VA initial values in boys were 91.8 ± 4.9%, 90.9 ± 5.6% and 94.7 ± 3.3% at SHORT, OPT and LONG, respectively. The corresponding values in men were 92.7 ± 4.0%, 92.6 ± 4.3% and 96.2 ± 2.6%. No significant difference was found for initial VA values between both groups and the three MTU lengths. VA decreased similarly in boys and men at SHORT (-17.4 ± 11.6 % and -14.7 ± 10.9%, respectively; Fig. 4A). VA decreased only in boys at OPT (-36.4 ± 20.3%; Fig. 4B) and to a greater extent than men at LONG (-20.4 ± 11.7% and -11.8 ± 5.8% in boys and men, respectively; Fig. 4C). *Post-hoc* tests also revealed that VA decrement was similar at SHORT than at LONG in men. However, boys showed a greater VA decrement at OPT than at SHORT (p < 0.001) and LONG (p < 0.001).

Overall, a significant correlation was found between relative VA changes and initial absolute torque level (R² = 0.18, p < 0.05): the stronger the individual, the lesser the VA reduction during the fatiguing exercise.

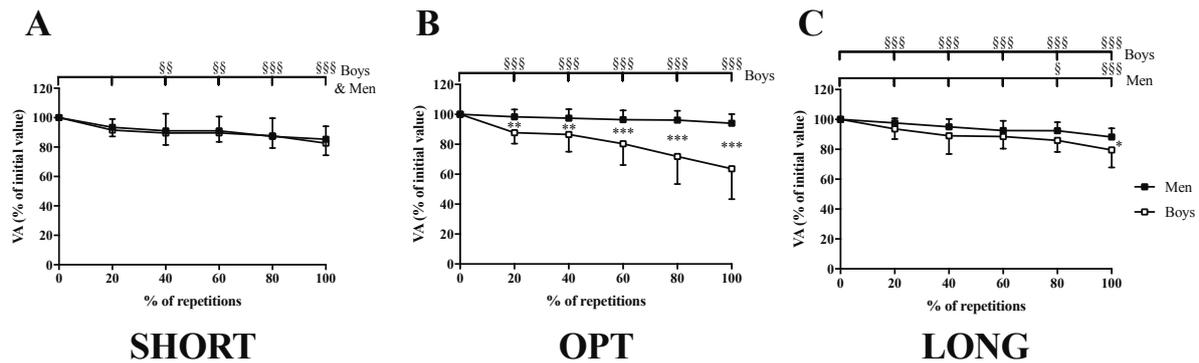

*Figure 4*: Time course of voluntary activation level (VA) at (A) short (SHORT), (B) optimal (OPT) and (C) long (LONG) MTU lengths during fatigue protocols in boys and men. §, §§ and §§§: significantly different from the initial value at $p < 0.05$, $p < 0.01$ and $p < 0.001$, respectively; *, ** and ***: significant difference between boys and men at $p < 0.05$, $p < 0.01$ and $p < 0.001$, respectively.

*Normalized EMG.* ANOVA also revealed a significant interaction effect (age group × MTU length × %REP) for absolute VL and RF RMS/$M_{max}$ values [$F(10;420) = 1.9$; $p < 0.05$, $\eta^2 = 0.05$, power = 0.86 and $F(10;420) = 2.2$; $p < 0.05$, $\eta^2 = 0.06$, power = 0.92, respectively]. VL and RF RMS/$M_{max}$ decreased at the three MTU lengths in boys, while men displayed a decrement of VL and RF RMS/$M_{max}$ only at SHORT. For absolute VM RMS/$M_{max}$ values, ANOVA revealed a significant age group × %REP interaction effect [$F(5;210) = 4.1$; $p < 0.01$, $\eta^2 = 0.10$, power = 0.95]. VM RMS/$M_{max}$ significantly decreased in boys whatever the MTU lengths investigated ($p < 0.001$), while no change was found in men.

*Co-activation level.* A significant interaction effect (age group × MTU length × %REP) was found for absolute %CoAct values [$F(10;420) = 2.3$; $p < 0.05$, $\eta^2 = 0.06$, power = 0.93]. %CoAct significantly decreased in boys at the three MTU lengths investigated ($p < 0.001$), whereas it only decreased in men at SHORT ($p < 0.05$). In addition, ANOVA showed a significant age group × MTU length interaction effect for %CoAct when expressed as percentage of its initial values [$F(2;84) = 5.9$; $p < 0.01$, $\eta^2 = 0.14$, power = 0.86]. However, *post hoc* tests revealed that relative decrement of the %CoAct was similar in boys and men at SHORT.

## DISCUSSION

The main purpose of the present study was to compare the development and etiology of neuromuscular fatigue between boys and men at different knee extensor MTU lengths. We hypothesized that at short MTU length (i.e. where differences in torque level between boys and men are reduced), both groups would display similar reduced amount of fatigue, mainly accounted for by central fatigue, rather than peripheral fatigue. The results of the present study partially confirm our hypothesis.

**Boys-men differences in performance fatigability**

As expected, the men achieved more repetitions at SHORT than at OPT or LONG. This result is consistent with previous studies in adults, reporting that performance fatigability is greater at long MTU length (20, 21). It has been suggested that the lesser performance fatigability at SHORT could be attributed to a lower voluntary muscle activation level (21). Indeed, a lower activation level implies greater resistance to fatigue (36). However, in the present study, no difference in the initial VA before fatigue was found between boys and men at any the MTU length. Thus, other factors should be considered.

Differences in torque level could explain differences in performance fatigability between boys and men. Indeed, at OPT, where the difference in torque level between boys and men is the greatest (223.9 N.m), children achieved more repetitions and as a result fatigued less rapidly than men. These results are consistent with a recent review (10), which suggests that the lower torque level in children than adults, would allow children to better resist to fatigue. Consistently, at SHORT, where the difference in torque level between boys and men was near twice lower than at OPT (123.3 vs. 223.9 N.m, respectively), both groups displayed similar performance fatigability. Thus, this lower difference in torque level at SHORT could reduce differences in performance fatigability between boys and men. This result is consistent with the statistical analysis performed by Ratel et al. (11), which shows that when the initial torque level is used as a co-variable, there are no further differences in performance fatigability between children and adults.

However, our results also show that performance fatigability was lower at LONG in boys than men, despite comparable differences in torque level to SHORT (136.5 vs. 123.3 N.m, respectively). Thus, the torque level cannot fully account for the differences in neuromuscular fatigue between boys and men. This suggests a specific effect of MTU length on neuromuscular fatigue being different between populations.

**Boys-men differences in the etiology of neuromuscular fatigue**

**Peripheral fatigue**

The lower performance fatigability in children is generally associated with a lower peripheral fatigue than adults (2, 3, 11). In the present study, peripheral fatigue was lower in boys than men at OPT, as evidenced by the lower reduction in $Qtw_{pot}$. This lower peripheral fatigue could be explained by a lower alteration in the contractile properties in boys than men, as illustrated by the lower decrement in $Dt_{100Hz}$. However, this finding should be interpreted with caution since $Dt_{100Hz}$ was evoked using a submaximal intensity to reduce discomfort. Although the use of a submaximal intensity may bias the evaluation of $Dt_{100Hz}$ (37), this limitation should nevertheless similarly affect children and adults. The $Dt_{10Hz}/Dt_{100Hz}$ decreased in both groups, no difference was identified between boys and men. Thus, the alteration in the E-C coupling cannot explain the differences in peripheral fatigue between boys and men. This result is not consistent with previous studies, which showed a greater alteration in the E-C coupling in men with the KE and PF muscles (8, 38). The lack of significant difference in the present study could be partly explained by small variations of this parameter especially at SHORT (22). In addition, the difference in peripheral fatigue between children and adults could not be explained by an alteration in sarcolemmal excitability, as evidenced by the lack of change in $M_{max}$ in any muscles investigated. This result is consistent with previous studies, reporting no change in $M_{max}$ during maximal intermittent contractions with the KE and PF muscles (8, 11) but inconsistent with other studies showing either a similar decrement during maximal sustained contraction (3) or an increase in children and a significant decrease in adults during dynamic contractions (39). Variations in fatigue exercise may explain these discrepancies

Boys and men exhibited a reduced peripheral fatigue during the fatigue protocol performed at SHORT. Indeed, $Qtw_{pot}$ slightly decreased in men and remained unchanged in boys. This result agrees with the fact that peripheral fatigue was found to be lower at short than at long MTU length in adults (21). The lower peripheral fatigue in these two groups could be attributed to the lower torque levels at SHORT, thereby limiting any alteration in contractile properties (21, 22). Accordingly, $Dt_{100Hz}$ and $M_{max}$ remained unchanged in boys and men after the fatigue protocol at SHORT. Taken together, peripheral fatigue at SHORT and OPT suggests that differences in torque level between children and adults may have a significant role in the peripheral mechanisms of neuromuscular fatigue. This interpretation was also supported by the significant inverse correlation between the initial absolute torque level and the relative $Qtw_{pot}$ changes.

However, torque level could not be the sole factor being affected when manipulating MTU length. Indeed, the results of the present study also highlighted that boys-men differences in peripheral fatigue were significant at LONG, even though the differences in torque level between boys and men were similar to SHORT. These discrepancies point out to a different specific effect of MTU length in boys and men. The greater peripheral fatigue in men at LONG than at SHORT could be attributed to potentially greater muscle damage. Indeed, previous studies showed that adults display significantly greater muscle damage when they perform isometric contractions at long than short MTU length (40). On the contrary, children could display less muscle damage whatever the MTU lengths. This specificity of children has already been reported in prepubertal boys by Chen et al. (41) following an eccentric exercise. This lower susceptibility of children to muscle damage could be attributed to their lower musculotendinous stiffness (42). Indeed, a compliant tendon could play a role of "mechanical buffer", thereby limiting muscular alterations (25). Nevertheless, these hypotheses cannot be tested from the data obtained in the current study.

A lower tendon stiffness may also reduce the energy cost of muscle contraction *in vitro* (25) and as result induce less performance fatigability. This could be particularly true in prepubertal children given that they have a lower musculotendinous stiffness than their older peers (42). Therefore, differences in energy cost between age groups could explain differences in peripheral fatigue between children and adults at LONG. However, although muscle oxygenation is an indirect measurement of energy cost, these data of the present study do not support this assumption. The alteration in muscle oxygenation was found to be lower in boys than men at any MTU length. Furthermore, MTU length did not have any effect on energy cost in both age groups. These findings are consistent with previous studies reporting no effect of MTU length on energy cost during voluntary contraction (43). Nevertheless, the effect of MTU length on energy cost is still debated in the literature, given that other studies have reported a lower energy cost at short MTU length (24, 44). In addition, the differences in muscle oxygenation between children and adults could be also attributed to different intramuscular pressures, related to the difference in torque level between boys and men. Indeed, a higher torque is generally associated with a higher intramuscular pressure and potentially a greater TSI decrease (45), as is observed in the present study in men compared to boys.

**Central fatigue**

Boys displayed a greater decrease in VA and $RMS/M_{max}$ of the VL, VM, and RF muscles than men when the fatigue protocol was performed at OPT. These results are consistent with previous studies reporting that children's neuromuscular fatigue is more from central origin during sustained or intermittent KE contractions (2, 8, 11). The decrement in agonist KE activation was also accompanied by a greater decrement of antagonist co-activation in boys. This neural modulation may have contributed to preserve the net joint torque production at the three MTU length during the fatiguing exercise in boys and consequently limited the development of peripheral fatigue within the agonist muscles in boys.

Consistently with our assumptions, differences in central fatigue between boys and men were reduced at SHORT, as evidenced by the comparable decrement of VA and $RMS/M_{max}$ of VL and RF between the two populations. These results are consistent with previous studies, which show that neuromuscular fatigue is mainly accounted for by central fatigue at short rather than at long MTU length in adults (20, 21). The authors suggested that at short MTU length, the torque level developed would be lower than at long MTU length and a lower torque level would induce a longer exercise duration until the same level of exhaustion and thus a greater central fatigue. There would be an interdependence between exercise duration and central fatigue; the longer the exercise duration, the greater the central fatigue (46). In the present study, men performed exercise for longer at SHORT, thereby potentially promoting greater central fatigue. The inverse seems to be observed in pre-pubertal boys; in fact, the VA decrement was lower at SHORT than at OPT, despite they performed exercise for longer at SHORT than at OPT. This result may argue for a specific MTU length effect being different between boys and men. Nevertheless, the similar and significant central fatigue in boys and men at SHORT seems to be explained by the reduced difference in torque level between these two groups (supported by the significant correlation between the initial absolute torque and relative VA changes), which has potentially promoted long exercise duration. In addition to the lower torque level, the decrement in %CoAct observed at SHORT in boys and men could have also limited the development of peripheral fatigue by preserving the net joint torque.

On the other hand, our assumptions are not confirmed by central fatigue at LONG. Indeed, boys displayed a greater central fatigue than men, as evidenced by greater decrement in VA and normalized EMG, despite the reduced difference in torque level. These results could be attributed to the children's specificity to central fatigue during maximal intermittent contractions (8) and/or to the fact that MTU length has a different specific effect between children and adults. However, the experimental setting of the present study does not allow identifying the factors underpinning differences in central fatigue between pre-pubertal children and adults at LONG. The effect of MTU length on central fatigue and its origins (spinal and supra-spinal) remains to be clarified in children and adults.

## CONCLUSION

MTU length has an effect on child-adult differences in the development and etiology of neuromuscular fatigue. At optimal and long MTU lengths, children display lower performance fatigability than adults, mainly accounted for by central fatigue, rather than peripheral fatigue.

In contrast, at short MTU length, the differences in neuromuscular fatigue between children and adults are significantly reduced. Both populations have a reduced performance fatigability, associated with a limited peripheral fatigue and a significant amount of central fatigue. These lower differences in fatigue at short MTU length could be explained by reduced differences in torque level between children and adults compared to optimal MTU length. Nevertheless, the difference in torque level could not account for all the results of the present study; other factors such as muscle damage, metabolic profile and specific neural regulations in children should be considered. Future studies are required to understand more extensively the effect of MTU length on the differences of neuromuscular fatigue between prepubertal children and adults.

## ACKNOWLEDGMENTS

The authors thank the participants for their time and effort.

No support was provided for this study by any manufacturer of the instruments used. Additionally, the results of the present study do not represent an endorsement of any product by the authors or the ACSM.

## CONFLICT OF INTEREST

No conflicts of interest, financial or otherwise, are declared by the authors. Further, the authors declare that the results of the study are presented clearly, honestly, and without fabrication, falsification, or inappropriate data manipulation.